\documentclass[preprint]{emulateapj-rtx4}
\pdfoutput=1
\usepackage{rotating}

\usepackage{graphicx}
\usepackage{epstopdf}
\usepackage{natbib}
\usepackage{xspace}
\usepackage{color}
\usepackage{amsmath}

\newcommand{\jykms}  {~Jy~beam$^{-1}\cdot$km~s$^{-1}$\xspace}
\newcommand{\kms}     {~km~s$^{-1}$\xspace}

\newcommand{\mjy}     {~mJy~beam$^{-1}$\xspace}
\newcommand{\mujy}   {~$\mu$Jy~beam$^{-1}$\xspace}

\newcommand{\msun}  {~$M_{\sun}$\xspace}
\newcommand{\mearth}{~$M_{\earth}$\xspace}

\newcommand{\co}       {$^{12}$CO\xspace}
\newcommand{\ceto}    {C$^{18}$O\xspace}

\begin{document}

\title{Strongly misaligned triple system in SR~24 revealed by ALMA }
\author{M. Fern\'andez-L\'opez\altaffilmark{1}}
\author{L.~A. Zapata\altaffilmark{2}}
\author{R. Gabbasov\altaffilmark{3}}

\altaffiltext{1}{Instituto Argentino de Radioastronom{\'{\i}}a (CONICET), CCT La Plata, 1894, Villa Elisa, Argentina; manferna@gmail.edu}
\altaffiltext{2}{Instituto de Radioastronom{\'{\i}}a y Astrof{\'{\i}}sica, Universidad Nacional
Aut\'onoma de M\'exico, P.O. Box 3-72, 58090, Morelia, Michoac\'an, M\'exico}
\altaffiltext{3}{Instituto de Ciencias B\'asicas e Ingenier\'{i}as,
Universidad Aut\'onoma del Estado de Hidalgo,
Ciudad Universitaria. Carr. Pachuca - Tulancingo Km. 4.5 s/n
Col. Carboneras, Mineral de la Reforma, Hidalgo, M\'exico, CP 42184}

\begin{abstract}
We report the detection of the 1.3~mm continuum and the molecular emission of the disks of the young triple system SR24 by analyzing ALMA (The Atacama Large Millimeter/Submillimter Array)  subarcsecond archival observations. We estimate the mass of the disks (0.025\msun and $4\times10^{-5}$\mearth for SR24S and SR24N, respectively) and the dynamical mass of the protostars (1.5\msun and 1.1\msun). A kinematic model of the SR24S disk to fit its \ceto~(2-1) emission allows us to develop an observational method to learn what is the tilt of a rotating and accreting disk. We derive the size, the inclination, the position angle and the sense of rotation of each disk, finding that they are strongly misaligned ($108\degr$) and possibly rotate in opposite directions as seen from Earth, in projection. We compare the ALMA observations with \co SMA archival observations, which are more sensitive to extended structures. We find three extended structures and estimate their masses: a molecular bridge joining the disks of the system, a molecular gas reservoir associated with SR24N and a gas streamer associated with SR24S. Finally we discuss on the possible origin of the misaligned SR24 system, pointing out that a closer inspection of the northern gas reservoir is needed to better understand it.
 
\end{abstract}


\section{Introduction}\label{intro}

Multiple stellar and protostellar systems are commonly found in the Galaxy \citep[e.g.,][and references therein]{2013Duchene,2016Tobin}. Given the low probability of the stellar encounter and capture scenario in the average stellar density Galaxy \citep{2005Thies,2014Salyk}, the main ways to form a multiple system are either via turbulent fragmentation of the molecular cloud \citep[e.g.,][]{2010Offner}, by fragmentation of an initially inhomogeneous rotating and infalling core \citep[e.g.,][]{2014Boss}, or by fragmentation of a protostellar disk through gravitational instability \citep[e.g.,][]{2010Vorobyov,2011Kratter,2009Stamatellos}. 

If the multiple system is formed in a rotating core or a fragmenting massive disk, aligned co-rotating disks are expected, while in the case of turbulent fragmentation a multiple system with misaligned disks can be formed \citep{2000Bate}. A misaligned system can also be formed in a more dense clustered environment, where gravitational interactions between its members or even captures of passing-by (proto)stars can become relevant. The infall of material with an odd angular momentum onto one of the protostellar disks of the multiple system may also produce a tilt on its inclination, leaving the system misaligned. Therefore, studying disk alignment can provide answers on how a young multiple system has formed. 

A number of these misaligned young multiple systems have been observed. Optical polarimetric observations \citep[e.g.,][]{2006Monin}, misaligned outflows \citep[e.g.,][]{2016Lee}, direct imaging of disks \citep{1998Stapelfeldt,2008Kang,2009Ratzka,2011Roccatagliata} and more recently, high angular resolution of disk's gas kinematics \citep{2014Jensen,2014Salyk,2014Williams} constitute the evidence for misaligned circumstellar disks in multiple systems. In addition, most of these systems are wide binaries (separations $\gtrsim100$~AU) suggesting the necessity of turbulent gas motions or gravitational interactions in their formation. In this contribution, we present ALMA millimeter observations toward the triple system SR24 showing that disks in this system are strongly misaligned and possibly counter-rotating in projection. 

SR24 is a hierarchical triple system situated in the outskirts of the Ophiuchus molecular cloud \citep[$d=120$~pc,][]{2008Loinard}, southwest of clump E of L1688 \citep{1949Struve}. The system comprises three T~Tauri protostars \citep[SR24S, SR24Nb and SR24Nc, following the notation by ][]{2006Correia} with estimated spectral types and masses of K2, K4-M4 and K7-M5, and $>1.4$, 0.6 and 0.3\msun, respectively \citep{2006Correia}. Their ages range $10^5-10^6$~yr \citep{2003Potter}. When convenient, we use SR24N to refer to the northern close binary system, which from 1991 to 2004 changed the projected separation between SR24Nb and SR24Nc from $0\farcs2$ to $0\farcs080$, rotating in a clockwise direction \citep[proper motions measurements along four epochs can be found in Fig. A.2 of the appendix A of][and see also references therein]{2006Correia}. Likewise, SR24N is separated from SR24S by $5\farcs2$ \citep{1993Reipurth1} in a north-south direction. Infrared excesses \citep{1994Greene}, H$\alpha$ emission and Li absorption \citep{1998Martin} and 10~$\mu$m silicate emission \citep{1998Nuernberger}, all found toward SR24N and SR24S, are clear signposts of the presence of circumstellar accretion disks. Moreover, HST and H-band polarization intensity images \citep{2003Potter}, along with a coronograph IR image of SR24 \citep{2010Mayama} showed the emission due to the scatter of the stellar light in the circumstellar dust and gas of the SR24N and SR24S disks, with measured radii of $2\farcs0$ and $2\farcs6$, respectively. At millimeter wavelengths, only the continuum emission from SR24S has been detected \citep{1993Reipurth2,1994Andre,1998Nuernberger,2017Pinilla}, suggesting that only the warm dust at the inner part of the SR24N disk is still present while lacking the cold circumstellar dust of the outer part. Recent high-angular resolution (sub)millimeter observations of the SR24S circumstellar disk \citep{2005Andrews,2008Patience,2009Isella,2010Andrews,2010Ricci,2017Pinilla} unveiled the presence of a $\sim0\farcs25$ cavity at its center, making it a good candidate for an insight study on transitional disks. Molecular CO observations also detected the emission from the SR24S disk \citep{2005Andrews,2008Patience,2009Isella}. The detected width of the CO~(2-1) disk line is $\sim2$\kms, indicating the presence of a moderately inclined disk \citep{2009Isella}. In addition, slightly north of the SR24N position, there is a CO molecular extended source which probably covers the molecular emission from the SR24N disk. Interestingly, this source is stronger than the CO emission coming from the SR24S position \citep{2005Andrews}. Other environment features have been reported via optical and infrared scattered light observations \citep{2003Potter,2010Mayama}: (i) a bridge between the two disks that expands and bifurcates into a filamentary structure and (ii) a proposed spiral arm southwest of the SR24S disk.

The paper is organized as follows. 
In Section 2, we describe the ALMA observations. In Section 3, we present the main results obtained from the ALMA observations. In Section 4, we discuss our findings. Finally, in Section 5, we make a summary of this work.

\section{Observations}
\subsection{ALMA archive observations}
The data analyzed here come from two observations taken in 
the ALMA Cycle 1 (projects 2013.1.000912.S and 2013.1.00498.S). These projects
focused on the study of the SR24S disk properties as inferred from the 1.3~mm
(227~GHz) continuum and several CO isotopologues. Here we explain the
geometry of the disks using the continuum emission and the 
$^{12}$CO~(2-1) and C$^{18}$O~(2-1)  lines. We also discuss the nature of the 
$^{12}$CO~(2-1) extended emission in the SR24 triple system.

We built a 1.3~mm continuum image and a C$^{18}$O~(2-1) velocity cube 
using the data collected on September 26th, 
2015 (project 2013.1.000912.S). The time on source was 23.3~minutes. At that time, 
ALMA had 34 operative antennas with
baselines ranging up to 2270~m. During the observations the 
median system temperature was about 70~K. The phase center was at 
$(\alpha,\delta)$J2000.0=($16^h26^m58\fs506$,$-24\degr45\arcmin35\farcs87$),
and the FWHM of the primary beam of the telescope was $27\arcsec$ at the 
observing frequency. The ALMA digital correlator was set to have two continuum 
windows (2000~MHz width) and two 
additional spectral windows (1875~MHz width) most of which where line-free and were hence
added to the continuum after removing the line channels. The four windows were centered 
at 219.550~GHz, 220.383~GHz, 234.507~GHz and 232.987~GHz. Observations of the asteroid
Pallas were used to set the absolute scale for the flux density calibration. The quasar J1517-2422
was used to correct for bandpass and the quasar J1617-2537, observed every 10 minutes, was used 
to extract the time-dependent gains and calibrate the phases and amplitudes. We self-calibrated and
cleaned the final continuum image three times and applied a mask surrounding the two main sources detected in SR24.
The data were calibrated, imaged and analyzed using CASA \citep{2007McMullin}, KARMA \citep{1996Gooch}, MIRIAD \citep{1995Sault} and GILDAS\footnote{GILDAS data reduction package is available at http://www.iram.fr/IRAMFR/GILDAS.}  
packages. The final 7.5~GHz continuum image was obtained using 
natural weighting. Its synthesized beam is $0\farcs18\times0\farcs12$ with a P.A.$=71.5\degr$ 
and its rms noise level is 44\mujy. The C$^{18}$O~(2-1) velocity cube was obtained after applying the selfcal solution. The synthesized beam of the C$^{18}$O data is $0\farcs21\times0\farcs16$ with a P.A.$=76.6\degr$ 
and a rms noise level of 1.2~mJy~beam$^{-1}$ per 1.35\kms channel.

The $^{12}$CO~(2-1) dataset was obtained on July 21st, 2015 (project 2013.1.00498.S). SR24 was observed with 44 ALMA 12~m antennas for 12.6~minutes, with 946 baselines ranging from 15.1~m to
1600~m. During the observations the median system temperature was about 95~K. The phase center was at $(\alpha,\delta)$J2000.0=($16^h26^m58\fs499$,$-24\degr45\arcmin37\farcs42$). The ALMA correlator setup included eight spectral windows ranging in frequency from 215~GHz to 233~GHz. One of these windows, centered at 230.714~GHz, covers 469~MHz and contains the $^{12}$CO~(2-1) transition ($\nu_{rest}=230.538$~GHz). This spectral window has 1920 channels, which provides a spectral resolution of 244~kHz, which results in 0.3\kms at the observing frequency.  
In this case, Saturn's moon Titan was used as the absolute flux calibrator and the quasars J1517-2422 and J1627-2426 were used to get the bandpass and the gains corrections, respectively. CASA, KARMA and Miriad were again used to calibrate, image and analyze the \co data. Two \co velocity cubes were obtained. The first velocity cube was obtained after applying a uv-taper to stress the extended \co emission. This image has a synthesized beam size of $0\farcs60\times0\farcs54$ with a P.A.$=89.2\degr$ and it has an rms noise level of 4.6\mjy in the binned channels of 0.6\kms. A second velocity cube was made with a natural weighting, which results in a $0\farcs25\times0\farcs21$ with a P.A.$=61.3\degr$ synthesized beam and an rms noise level of 5.2\mjy per 0.6\kms channel.

\begin{deluxetable*}{ccccccccc}
\tablewidth{0pt}
\tablecolumns{9}
\tablecaption{Characteristics of the 1.3~mm continuum sources}
\phs
\tablehead{
\colhead{Source} & \colhead{Right Ascension} & \colhead{Declination} & \colhead{Flux Density} & \colhead{Semiajor Axis\tablenotemark{(a)}} & \colhead{Semiminor Axis\tablenotemark{(a)}} & \colhead{P.A.} & \colhead{i\tablenotemark{(b)}} & \colhead{M$_{disk}$\tablenotemark{(c)}}\\
\colhead{}  & \colhead{J2000} & \colhead{J2000} & \colhead{(mJy)}  & \colhead{($\arcsec$)} & \colhead{($\arcsec$)} & \colhead{($\degr$)} & \colhead{($\degr$)} & \colhead{(M$_{\sun}$)}
}
\startdata
SR24S  & $16^h26^m58\fs504$ & $-24\degr45\arcmin37\farcs21$ & 211$\pm$4 & 0.70$\pm$0.06 & 0.50$\pm$0.06  & 212$\pm$3 & 44$\pm$6 & 0.025$\pm0.001$ \\
SR24N & $16^h26^m58\fs434$ & $-24\degr45\arcmin32\farcs24$ & 0.34$\pm$0.07 & \nodata\tablenotemark{(d)} & \nodata & \nodata & \nodata & 4$\times$10$^{-5}\pm$1$\times$10$^{-5}$ \\
\enddata 
\tablenotetext{(a)}{Deconvolved semimajor and semiminor axis of the ring-shaped disk.}
\tablenotetext{(b)}{Estimated from the major and minor axis lengths as the $\arccos(R_{min}/R_{maj})$.}
\tablenotetext{(c)}{The disk mass is estimated assuming a typical gas-to-dust ratio of 100.}
\tablenotetext{(d)}{This disk is unresolved with the present ALMA resolution.}

\label{Tcont}
\end{deluxetable*}

\subsection{SMA archive observations}
The observations were obtained with the Submillimeter Array\footnote{The Submillimeter Array (SMA) 
is a joint project between the Smithsonian Astrophysical Observatory and the Academia Sinica Institute 
of Astronomy and Astrophysics, and is funded by the Smithsonian Institution and the Academia Sinica.} 
during 2004 August 2, when the array was in its compact configuration.  
The data were taken with 7 antennas and double-sideband receivers with an 
IF frequency of 225.494~GHz.  The zenith opacity ($\tau$ at 230~GHz) measured with 
the NRAO tipping radiometer located at the Caltech Submillimeter Observatory was $\sim$0.1, 
which indicates good weather conditions. The $^{12}$CO(2-1) millimeter line at a frequency 
of 230.538~GHz was centered in the upper sideband.  The digital correlator of the SMA was 
configured with 2~GHz bandwidth in each sideband and
a spectral resolution of 1.06~km$^{-1}$ ($\sim$ 0.8~MHz) per channel. 
The phase center was at $(\alpha,\delta)$J2000.0=($16^h26^m58\fs5$,$-24\degr45\arcmin36\farcs67$.  Both objects SR24S and SR24N fall well inside of the SMA 
FWHM ($57\arcsec$) for this wavelength. The total on-source integration time for SR 24 was 1.7~h.

Observations of Uranus provided the absolute scale for the flux density calibration. 
The gain calibrators were the quasars 
J1743$-$038 and J1733$-$130, while Callisto was used for bandpass calibration.
The uncertainty in the flux scale is estimated to be between 15\% and 20\%, based on the SMA monitoring of quasars.

The data were analyzed in the standard manner using the IDL-based MIR software package\footnote{http://cfa-www.harvard.edu/$\sim$cqi/mircook.html} and
MIRIAD. We used the ROBUST parameter of CLEAN set to 2, to obtain a better sensitivity 
losing some resolution. The resulting line-image rms was 40~mJy~beam$^{-1}$ per channel at 
an angular resolution of $2\farcs0\times1\farcs5$ with PA = $-42.4\degr$.

\section{Results}
\subsection{Continuum emission from the disks}
\label{Scont}
\begin{figure*}[!h]
\epsscale{1}
\plottwo{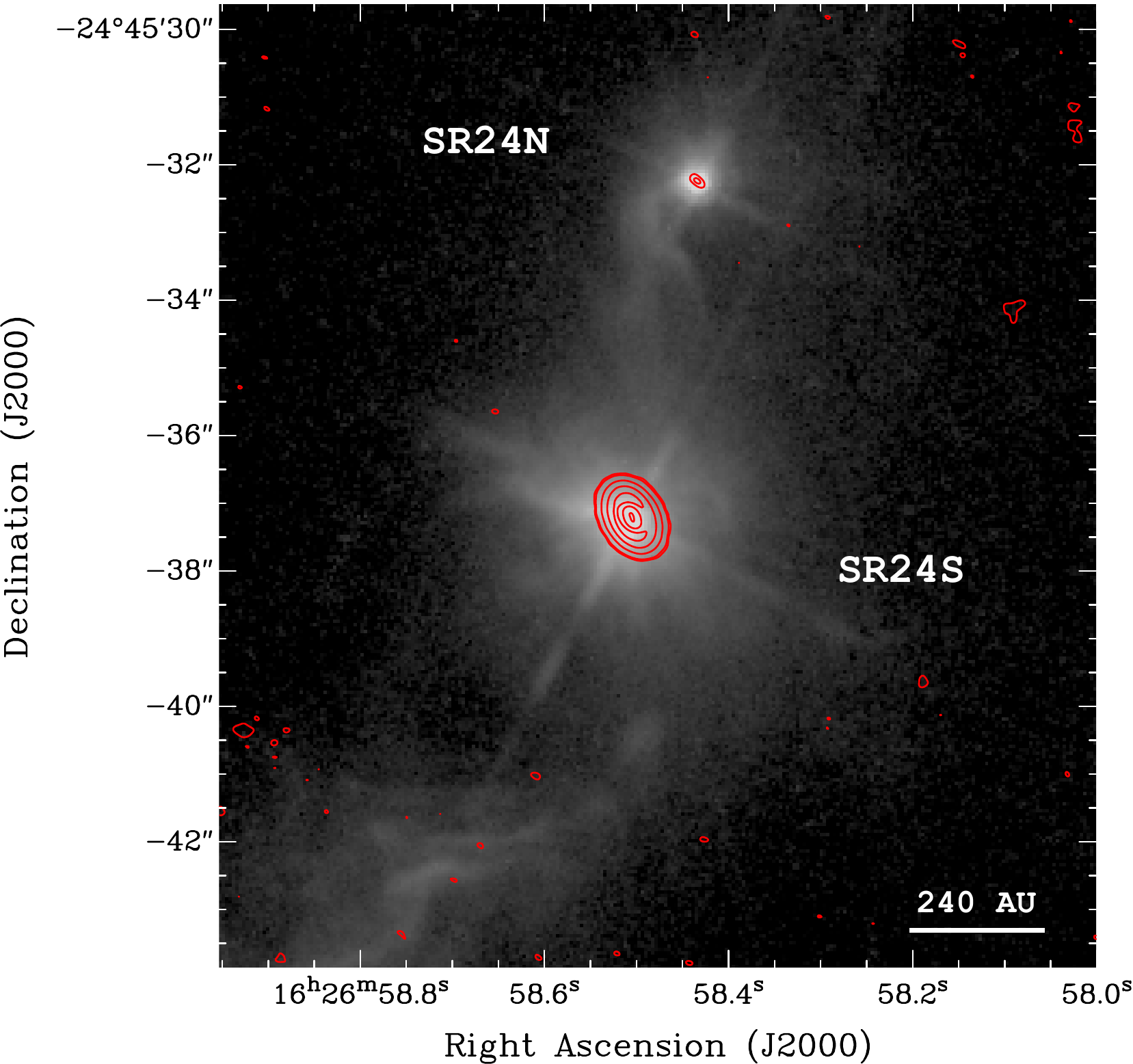}{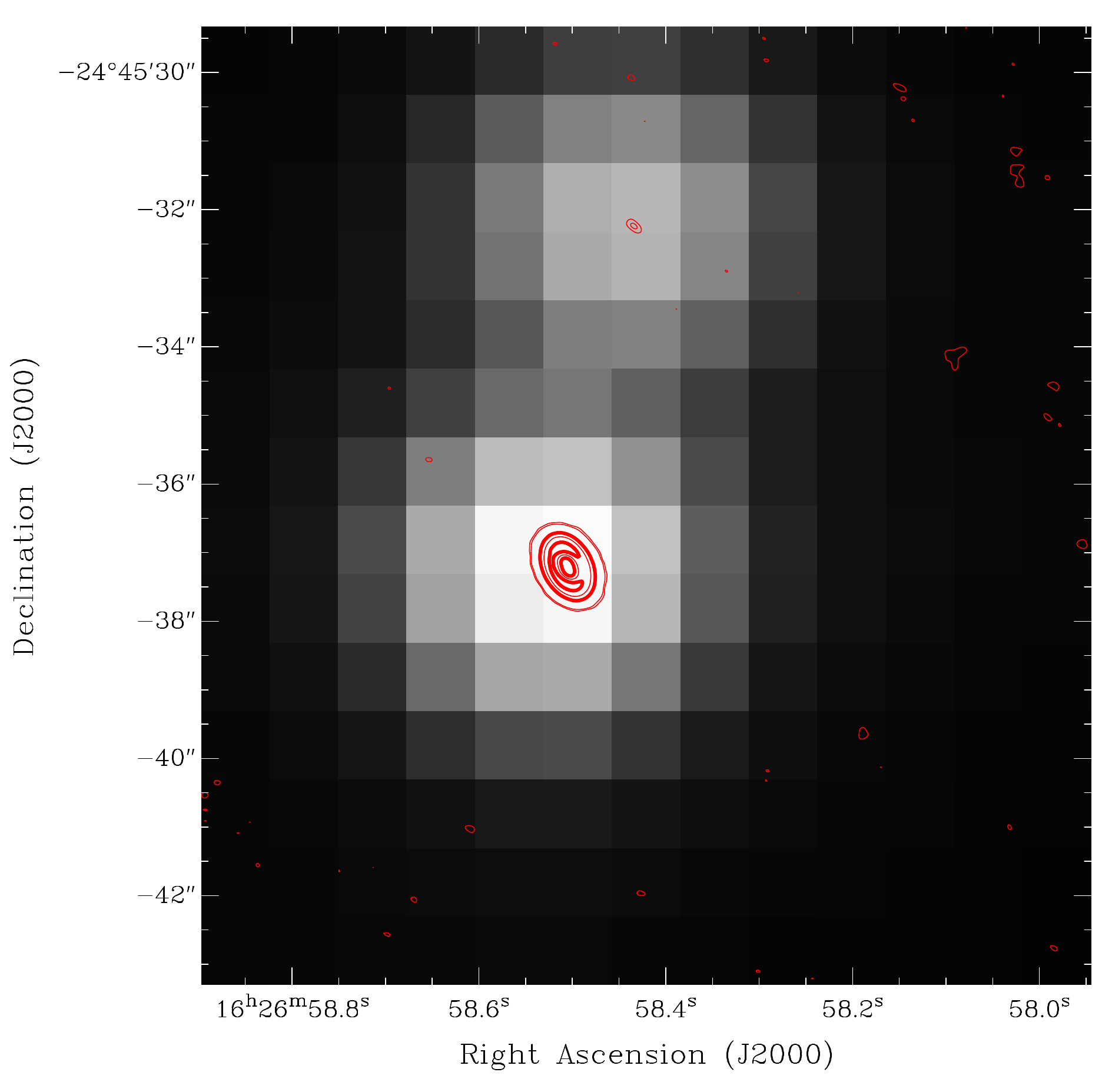}
\caption{\textbf{Left:} 1.3mm ALMA continuum image (red contours) on top of a f606w HST archive image (grey scale). Contours are 0.15,0.30,4,7,10\mjy. Rms noise level is 0.044\mjy. The synthesized beam is $0\farcs18\times0\farcs12$; P.A.$=71.5\degr$. \textbf{Right:} Same as for the left panel but with a 2MASS J-band image instead of the HST image. Note the good positional agreement (the astrometry uncertainty for the 2MASS data is about $0\farcs5$) of the two disks seen at IR (grey scale) and mm wavelengths (red contours).}
\label{Fcont}
\end{figure*}

The left panel of Figure \ref{Fcont} shows the spatial distribution of the ALMA 1.3~mm continuum emission
contours overlapped with an archival HST ACS $6000$\AA~image\footnote{The astrometry of the downloaded HST image was wrong, so we applied an offset to the center of that image in order to match the position of the SR24N disk with its optical counterpart.}. As previously reported, the system 
SR24 consists of two circumstellar disks (SR24N and SR24S) imaged 
for instance at infrared wavelengths \citep{2010Mayama}. While SR24S is associated with a well known
millimeter continuum source \citep[e.g.,][]{2005Andrews,2017Pinilla}, here we report for the first time the detection of SR24N in this wavelength domain. ALMA shows SR24S having a well-defined annular morphology. This ring of
dust is seen as an annular ellipse in projection (see Table \ref{Tcont}), with two brighter parts 
toward the northwest and the southeast. SR24N is unresolved with the ALMA 
resolution ($\sim0\farcs15$) and its position coincides with the 
position of the 2MASS disk within the positional uncertainty ($0\farcs5$, see right panel of Figure \ref{Fcont}). 
We measured a peak flux of $314$\mujy (that is a $7\sigma$ detection) for the SR24N millimeter source, which 
has a size of the order of the angular separation of the binary contained in it \citep[as measured by][]{2006Correia}.
For comparison, an M or K spectral type star at the distance of the Ophiuchus molecular cloud would have a flux density\footnote{Assuming an effective temperature of T$=3700$~K and a radius of R$=0.8$~R$_{\sun}$ it is possible to derive the flux density at $\nu=230$~GHz for a typical M or K star in the Ophiuchus cloud (d$=120$~pc) using 
$$S_{\nu}=\int_{source}I_{\nu}(\theta,\phi)\cos{\theta}d\Omega\quad.$$ For small circular sources $S_\nu\approx\pi\theta^2 I_\nu$, where $\theta$ is the angular size of the star and I$_\nu$ is the specific intensity. In this case 
$$\theta=\arcsin({R/d})\approx1.51\times10^{-10}~rad$$ and, using the Rayleigh-Jeans approximation, 
$$I_\nu=B_\nu\approx2kT\nu^2/c^2=6.0\times10^{-11}~erg\cdot cm^{-2}\cdot s^{-1}\cdot Hz^{-1}\cdot sr^{-1}\quad.$$ With all these values the derived flux density is 0.43$\mu$~Jy.} of $\lesssim0.5$\mujy.

Therefore, the detected
millimeter source in SR24N is probably due to the dust emission from the protostellar disk(s).
Assuming optically thin isothermal dust emission, a gas-to-dust ratio of 100, a dust opacity of 
1.1~cm$^2$~g$^{-1}$ at 1.3~mm appropriate for dust with thin ice mantles \citep{1994Ossenkopf},
an opacity spectral index with an exponent $\beta=0.4$ for both sources \citep{2009Isella,2010Ricci,2014Pinilla}, and a dust temperature of 30~K, we estimate a mass associated with the SR24N millimeter source
of $4\times10^{-5}$\msun or 13$\pm7$\mearth (where we have estimated the statistical error based on the uncertainty of the flux density and a 5~K uncertainty for the temperature). Using the same 
approximate values, for SR24S we obtain a mass of 0.025$\pm0.007$\msun in good agreement 
with previous estimates \citep[e.g., ][]{2009Isella,2010Andrews,2011Andrews}.

\subsection{$^{12}$CO Molecular emission from the disks}
\begin{figure*}[!h]
\epsscale{1.17}
\plotone{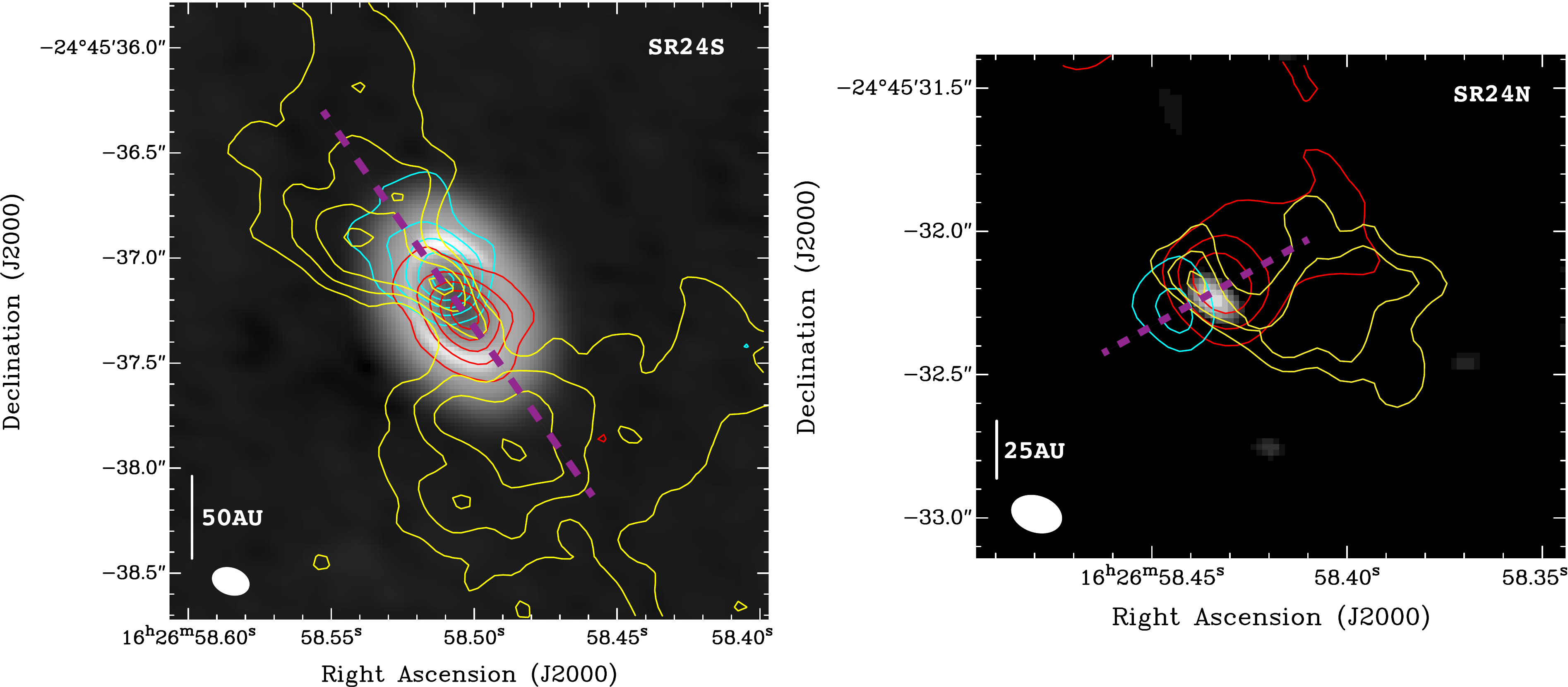}
\caption{\textbf{Left:} $^{12}$CO integrated emission from SR24S. Redshifted emission  is integrated from 7.3\kms to 12.4\kms (red contours at 27\%, 42\%, 69\% and 96\% of the peak emission); near zero-velocity emission  is integrated from 2.2\kms to 4.1\kms (yellow contours at 50\%, 65\%, 80\% and 95\% of the peak emission); blueshifted emission is integrated from -6.0\kms to 1.6\kms (blue contours, same as red-shifted contours). Contours are overlapped on top of a 1.3~mm continuum emission (grey scale) and the beam is shown in the bottom left corner. \textbf{Right:} Same display as in left panel for SR24N. In this case, redshifted emission is integrated from 6.6\kms to 10.5\kms (red contours at 20\%, 30\% and 40\% of the map peak emission located at SR24S); zero-velocity emission is integrated from 5.3\kms to 6.0\kms (yellow contours at 50\% and 60\% of the map peak emission); blueshifted emission is integrated from -0.3\kms to 2.2\kms (blue contours at 50\% and 60\% of the map peak emission).
}
\label{Fmoms0}
\end{figure*}

The \co molecular emission observed by ALMA in the SR24 system shows the kinematics of 
two gaseous rotating disks at the positions of SR24S and SR24N (Figure \ref{Fmoms0}). The gas content and kinematics of the SR24N disk is reported here for the first time\footnote{Note that \cite{2005Andrews} reported the detection of the SR24N disk in \co with the SMA, but a closer inspection on the coordinates of this source make it clear that the SMA emission stems from an extended source north of SR24N position.}. Table \ref{T12co} presents a summary of the properties of the disks found through the \co emission. 

\begin{figure*}[!h]
\epsscale{1}
\plotone{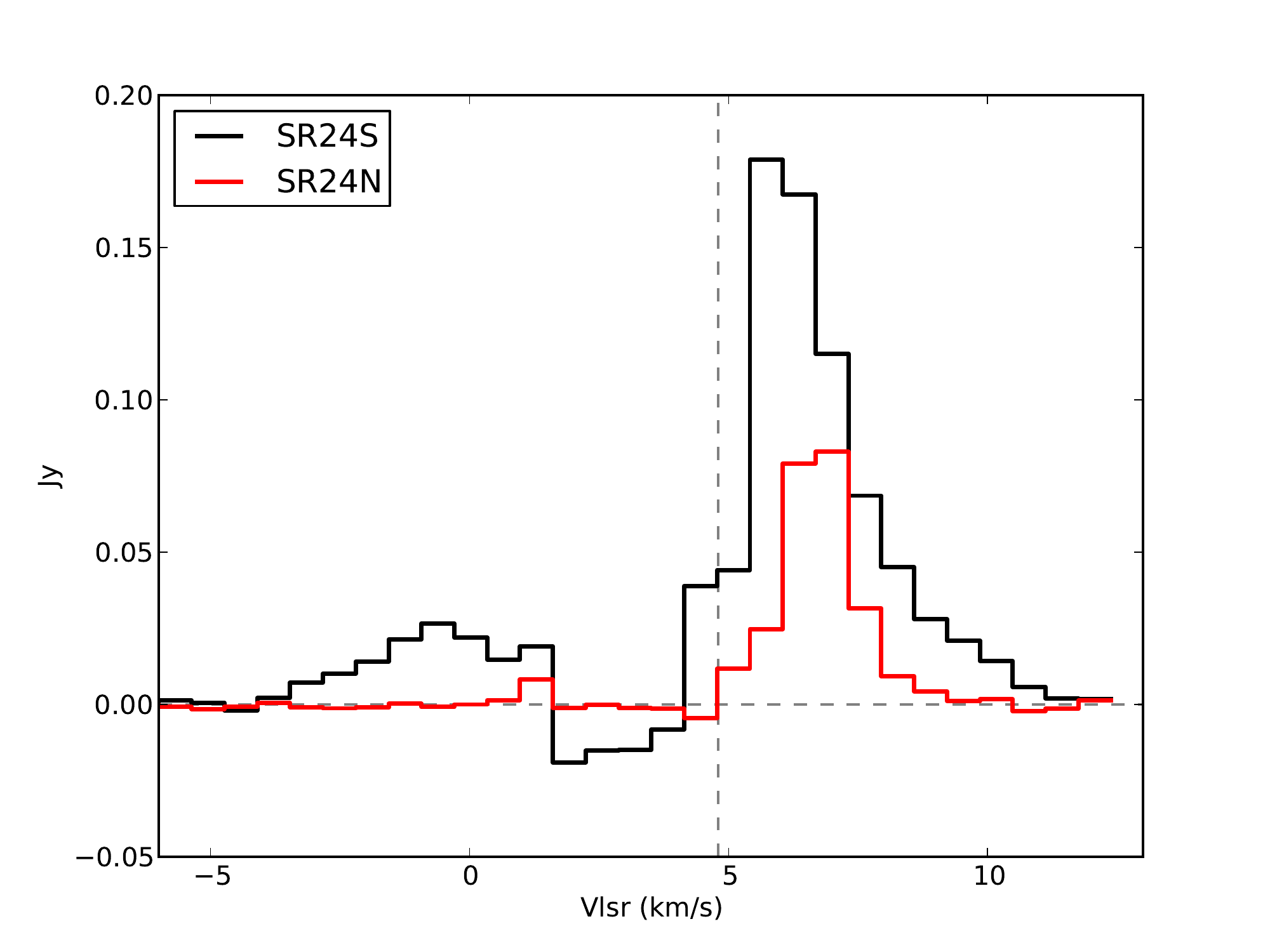}
\caption{\co integrated spectra over the two SR24 disks. The foreground cloud velocity at 4.8\kms is marked by a vertical line. The spectra show a skewed asymmetric double-peak profile with the stronger red-shifted peak, possibly due to foreground absorption.}
\label{Fspec}
\end{figure*}

A southwest-northeast (red- to blue-shifted emission, respectively) velocity gradient indicative of rotation is observed in SR24S at a position angle\footnote{Throughout this work we have adopted the convention typically used in visual binary orbit determination \citep{1978Heintz,2014Jensen}. The position angle (P.A.) is counted in degrees from north (0$\degr$) through east (90$\degr$), ranging from 0$\degr$ to 360$\degr$, pointing to the red-shifted edge of the disk (\textit{ascending node}). The inclination of the disk (\textit{orbit}) with respect to the plane of the sky ranges from 0$\degr$ to 180$\degr$, being 90$\degr$ that of an edge-on disk. If the disk rotates clockwise the inclination is $<90\degr$; if it rotates counterclockwise the inclination is $>90\degr$.}, P.A.$=218\degr\pm2\degr$, ranging for about 16\kms. The disk rotation is centered at 4.4\kms and the spectrum integrated over the disk shows a double peak, being stronger the red-shifted one (Figure \ref{Fspec}). In addition, four slightly blue-shifted spectral channels close to the cloud velocity (which we found to be the channel at V$_{cloud}=4.8$\kms, because it shows the typical interferometric pattern when extended emission is missed) show a significant absorption over the SR24S ring disk. This radial velocity range coincides exactly with the kinematics of the cloud 33 of the $\rho$-Ophiuchus region in which the SR24 system is embedded \citep{1990deGeus}. This absorption is possibly affecting the results of our 2D-Gaussian fit, which would explain the different inclination obtained, $70\degr$, with respect to that obtained by analyzing the continuum emission, $44\degr$.

\begin{figure*}[!h]
\epsscale{1}
\plotone{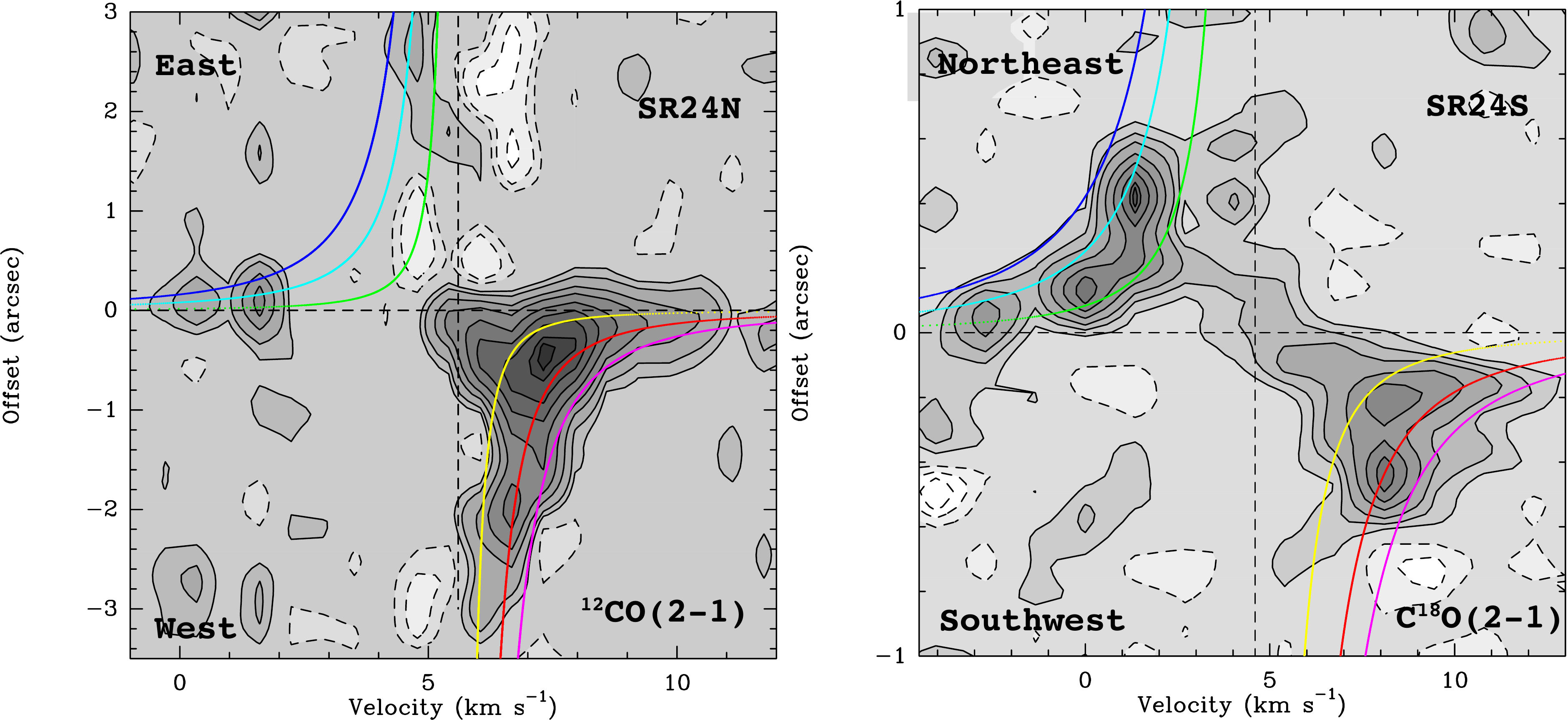}
\caption{\textbf{Left:} Position velocity diagram built from the \co emission along a cut with P.A.$=117\degr$ centered on SR24N. Contours are at -30,-20,-10,-5,5,10,20,30,45,60,75,90 and 96\% of the maximum peak emission. Keplerian curve profiles following the expression $v_{obs}=(GM_*sin^2(i)/r)^{1/2}$ are overlapped for different masses: green/yellow, 0.1\msun; cyan/red, 0.5\msun; blue/magenta, 1.0\msun. Dashed lines are drawn at the center of the disk in position and velocity (5.6\kms). 
\textbf{Right:} Same as in the left panel but showing a velocity cut along the major axis of the C$^{18}$O~(2-1) emission of the SR24S disk (P.A.$=32\degr$) and located at its central position. The masses for the overlapped Keplerian curves are: green/yellow, 0.5\msun; cyan/red, 1.5\msun;  blue/magenta, 2.5\msun. The central velocity in this case is 4.6\kms.
}
\label{Fpvs}
\end{figure*}

Centered at a different velocity (5.6\kms, that is 1.2\kms away from that of SR24S), the fainter SR24N \co emission presents a velocity gradient. The red- to blue-shifted emission gradient (west-east, respectively; Figure \ref{Fspec}) goes along the P.A.$=297\degr\pm5\degr$ with a velocity range of about 11\kms. The integrated spectrum over the northern disk also presents a skewed double-peaked profile, with a stronger red-shifted peak most probably due to the cloud absorption as in SR24S. Since SR24N disk is not detected in less optically thick molecular tracers we used the available \co information to make an order-zero estimate of the dynamical mass of the system. Left panel of Figure \ref{Fpvs} shows a position-velocity diagram taken along the major axis of the SR24N disk. For comparison we overlapped three Keplerian rotation curves produced with central masses ranging between 0.1\msun and 1\msun, which encompass the emission from the disk. Another crude estimate of the dynamical mass for SR24N can be made by using the expression $(M/M_{\sun})=0.00121\times((\Delta v/km~s^{-1})/2)^2\times(r/AU)$, where the parameters are $\Delta v=V_{range}$ and $r$ is equal to the geometric mean of the sizes given in Table \ref{T12co}. The resulting mass, 0.8\msun, is in good agreement with the total mass of the close binary system \citep[0.95\msun, ][]{2006Correia}. Compared with it, the mass of the SR24N disk is negligible.


\subsection{Simple modeling of the SR24S disk kinematics}
\label{Smodel}
The \ceto~(2-1) line emission is only detected toward SR24S and not SR24N. 
Although fainter than the \co line its emission is more optically thin and is less affected 
by the cloud absorption \citep{2017Pinilla}. This is the reason why the \ceto~(2-1) line 
is the best molecular proxy available to study the kinematics of the SR24S disk. We made a 
simple model of a rotating and accreting thin disk, which does not
include radiative transfer since the goal is to study the disk kinematics and geometry.
The surface density is described by a tapered power law consistent with the theoretical predictions 
for accretion disks \citep{1974Lynden,1998Hartmann,2017Facchini,2017Factor}:
$$\Sigma_{gas} (r) =\Sigma_0 \left(\frac{r}{r_c}\right)^{-\gamma}\exp{\left[-\left(\frac{r}{r_c}\right)^{(2-\gamma)}\right]},$$
where $\Sigma_c$ is proportional to the disk mass, $r_c$ informs about the radial extent of the gas disk and $\gamma$ is the power law index that we fixed to 0.8 following \cite{2010Andrews}. We implemented the kinematics of the rotating and accreting disk using the Miriad tasks \textit{velmodel} and \textit{velimage} and convolved the model with a 2D-Gaussian to match the observed synthesized beam. To build up the velocity cube with \textit{velimage} we assumed that both the rotation and the infall velocity have an $r^{-0.5}$ radial dependence.  Our model of the velocity cube has 9 free parameters: peak intensity, disk center $(x_0, y_0)$, inclination with respect to the line of sight $i$, the characteristic radius $r_c$, the position angle $P.A.$ (measured from North to East), the LSR velocity $v_{LSR}$,  the rotation velocity $v_{rot}$ and the accretion velocity $v_{infall}$ at a fiducial radius of $1\arcsec$ (120~AU for the adopted distance). We used an optimization algorithm based on the AGA method developed by \cite{2009Canto} to fit the observed velocity cube and ran it a 100 times in order to extract a standard deviation for the averaged parameters (Table \ref{Tc18o}).

\begin{figure*}[!h]
\epsscale{1}
\plotone{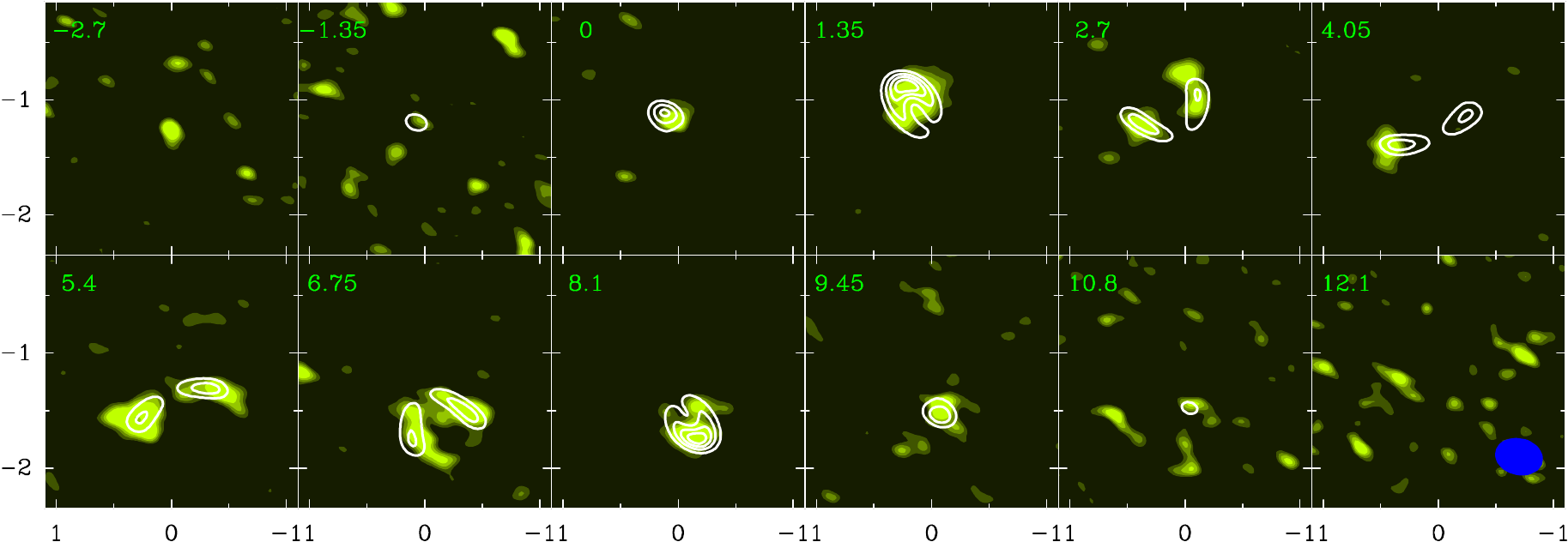}
\caption{Velocity cube of the ALMA \ceto~(2-1) emission of the SR24S disk (color scale) overlapped with a simple rotating and accreting disk model (white contours, see text). Color tones range from 20\% to 90\% of the peak flux (10.3\mjy\kms) in steps of 10\%, while contours are 3, 6, 9, 12, 15 times the rms noise level of the observed emission. The $v_{LSR}$ of every velocity channel is indicated at the top right corner and the synthesized beam of the observations ($0\farcs21\times0\farcs16$, P.A.$=77\degr$) is shown at the bottom right corner of the last velocity channel.}
\label{Ffit}
\end{figure*}

Figure \ref{Ffit} shows the good agreement between the simple model and the \ceto~(2-1) observed velocity cube, which displays the typical butterfly-shaped emission along the velocity channels, proper of a Keplerian disk. The blueshifted emission is located preferentially toward NE while the redshifted emission is found toward SW, with the velocity center estimated at 4.6$\pm0.2$\kms. The inclination and position angle match the values found in Section \ref{Scont} from the continuum emission, while the radius of the gaseous disk seems to be larger by almost a factor of 2 than the radius of the dusty disk, which may be explained by differences in the optical depth of the CO line and the continuum emission \citep{2017Facchini}. The rest of the resulting parameters of the fit are summarized in Table \ref{Tc18o}. The model included the rotation velocity of the disk at a fiducial radius of 120~AU ($v_{rot}=3.2\pm0.5$\kms), which we use to estimate the dynamical mass of the protostar ($M=v_{rot}^2 r/G$) of  1.5$\pm$0.5\msun. This mass makes a good fit to the position-velocity diagram taken along the disk's major axis (at P.A.$=212\degr$) shown in the right panel of Figure \ref{Fpvs}. It is, in addition, similar to the value of 2\msun reported by \cite{2010Andrews}.

\subsection{Molecular extended emission}
The \co~(2-1) ALMA and SMA velocity cubes show several regions of extended emission. The SMA dataset, has shorter baselines than the ALMA dataset and therefore is more sensitive to extended emission, although it has on the contrary, a coarser spectral resolution. We present both velocity cubes overlapped in Figure \ref{Fcube}. There are three main features that we point out: (i) a gas reservoir extending north-northwest of SR24N, (ii) a bridge of gas connecting SR24N with SR24S disks and (iii) an elongated and blueshifted feature due southwest of SR24S. 

The northern gas reservoir spreads out $7\farcs5$ east-west and extends for about $4\arcsec-5\arcsec$ north of SR24N. It embraces the SR24N disk and is probably associated with it, since it is detected from 5.4\kms to 7.9\kms, part of the red-shifted velocities at which the SR24N disk is seen. The elongated \co bridge connecting SR24N and SR24S disks extends for $\sim5\farcs5$ north-south. It is mainly detected between 5.4\kms and 6.0\kms with the SMA, although the ALMA observations also show a faint arc-like structure linking the northeast part of the SR24S ring-like disk with the east edge of the SR24N disk and the northern gas reservoir. There is another \co linearly elongated source due southwest from SR24S which is detected at blue-shifted velocities ranging from -0.9\kms to 0.9\kms. 

We estimate the mass of the northern gas reservoir and the gas bridge between the disks using the \co (2-1) SMA emission, assuming that the gas is in local thermodynamic equilibrium and its emission is optically thin. We derive them with the following expression:
$$M/M_{\sun}=1.03723\times10^{-4}~\frac{Q(T_{ex})~e^{(E_u/T_{ex})}~d^2\int{S_{\nu}dv}}{\nu^3~S\mu^2~(X_{CO}/X_{H_2})},$$
in which we adopt an excitation temperature $T_{ex}=30$~K, the partition function for the \co $Q(30K)=1.29$, the energy of the upper level of the \co~(2-1) transition is $E_u=16.596$~K, the fractional abundance $X_{CO}/X_{H_2}=10^{-4}$, the distance is d=0.12~kpc, the rest frequency $\nu=230.538$~GHz, the molecular dipolar momentum $S\mu^2=0.02423$~D and the integrated flux density $\int{S_{\nu}dv}$ goes in Jy~km~s$^{-1}$. The resulting masses, 13$\times10^{-7}$\msun and 2$\times10^{-7}$\msun (see Table \ref{Textended}), for the northern reservoir and the bridge respectively, are of the order of half the mass of the planet Earth.

\section{Discussion}
\subsection{Geometry of the disks}
Having measured the P.A. and the inclination of a disk, the only information that is missing to completely characterize its geometry is to know the \textit{tilt} of the protostellar disk, (i.e., which end of the disk's minor axis is the nearest to the observer). This could be addressed indirectly by putting together the information about the approaching-receding side of the disk and the sense of the disk's rotation \citep[an analogous problem to distinguish between leading and trailing spiral arms in galaxies; e.g.,][]{1958deVaucouleurs,2017Repetto}.

For SR24S the southwest side of the disk is receding (red-shifted), while the northeast is approaching (blue-shifted). We can now derive the sense of rotation (and hence the tilt of the disk) in three ways. The first, by admitting the indications of previous coronograph IR observations \citep{2010Mayama}, which show an extended arc of emission reported as a spiral arm of the SR24S disk. The curvature of the spiral arm would indicate that the disk is rotating counterclockwise (as seen in projection), which implies that east is its near side. Second, the same infrared observations show brighter emission toward the east side of the SR24S disk, which lead \cite{2010Mayama} to conclude that the near side is the east side (the same applies also for the SR24N disk). Along with the approaching/receding kinematic information, this implies a counterclockwise rotation for the SR24S disk. The third evidence supporting this sense of rotation comes from a close inspection of the velocity field of the \ceto disk. Figure \ref{Fpvperp} presents two different velocity cuts perpendicular to the major axis of the SR24S disk. A subtle asymmetric pattern can be noticed in both position-velocity diagrams. On one hand, the northern pv-cut shows the blueshifted emission in a boomerang-shaped feature originated by the rotation, but the west arm of the boomerang is shorter than the east arm. On the other hand, the southern pv-cut shows the redshifted emission in another boomerang-shaped feature, but this time the east arm is slightly shorter and fainter. The asymmetric length of the boomerang arms in both pv-diagrams is due to the non-zero infall velocity as supported by the following study that we present using some synthetic models. Actually the asymmetry is more pronounced as the infall velocity is larger and the disk is closer to edge-on. Note that if outflow rather than infall motions are considered, the asymmetry would be reversed. In the case of the SR24 system we assume infall motions based on the evidence of accretion provided by \textit{Paschen} hydrogen recombination lines observations by \cite{2006Natta}.

In Figure \ref{Fpvmodels} we show sketches for two rotating and accreting disks with inclinations separated by $180\degr$ and a third rotating disk with no accretion (Cases I, II and III, respectively). We simulated the kinematics of these disks using the toy model described in section \ref{Smodel} for SR24S, but changing the inclination and the accretion velocity while keeping the receding and approaching sides of the observations fixed. Two position-velocity cuts perpendicular to the major axis of the disks are presented for each case. Case I and III disks have their nearest side to the east (inclination $i=46\degr$) and differ only in the accretion velocity: $v_{I}=3.2$\kms and $v_{III}=0.0$\kms. Case II disk has $i=136\degr$, $v_{II}=3.2$\kms and its nearest side is west. Case I and II disks rotate in opposite directions (as seen in projection) so that they keep the receding and approaching sides the same. The pv-cuts perpendicular to the major axis of the disks show similar boomerang-shaped features as the SR24S observations. The boomerang shapes appear due to the gas rotation (Case III), but the asymmetries in the length of the boomerang arms are related to the accretion or infall motions (Cases I and II). Indeed, in the case of an accreting disk, the asymmetry pattern is also related with the inclination of the disk with respect to the plane of the sky. As evidenced by Figured \ref{Fpvmodels} when $i<90\degr$ (Case I) the infall motions make the nearest blueshifted quarter of the disk to reach less blueshifted velocities while the opposite applies for the farthest blueshifted quarter of the disk. Analogously, the farthest redshifted quarter appears less redshifted than the nearest redshifted quarter. On the contrary, when $i>90\degr$ (Case II) the asymmetries are just reversed, making it possible from the kinematic observations to determine which is the nearest side, the true inclination and the sense of rotation of the disk, given that the disk has infall motions. All this analysis leads us to derive the sense of rotation (counterclockwise) and the location of the near edge (east edge) of the SR24S disk.

For the SR24N disk, the lower signal-to-noise ratio and the foreground cloud absorption of the \co data prevent a similar analysis. Therefore, we assume that the sense of the disk rotation matches that of the close binary system SR24Nb and SR24Nc \citep[i.e., clockwise, see section \S\ref{intro} and ][]{2006Correia}. Although there are examples of disks hosting counter-rotating stars \citep[][and references therein]{2010Bate,2016Vorobyov}, the simplest hypothesis is that both binary and disk rotate in the same sense. 

Figure \ref{Fcartoon} sketches the SR24 system (note that it is not to scale and the gaps of the disks are not real), with the binary system SR24N rotating clockwise and with an inclination $<90\degr$, while the SR24S disk is rotating counterclockwise with an inclination $>90\degr$. Given the measured position angles and the inclinations of both disks (Tables \ref{Tcont}, \ref{T12co} and \ref{Tc18o}), we derive an angle between the angular momentum vectors of the two disks\footnote{We follow the expression given in \citep{2014Jensen}: \\$\cos{\Delta}=\cos{i_1}\cdot\cos{i_2}+\sin{i_1}\cdot\sin{i_2}\cdot\cos{(PA_1-PA_2)}$, \\where $i_1$ and $i_2$ are the disk inclinations.} of $108\degr\pm25\degr$. Even considering that the SR24N disk rotates in the opposite direction (it would have $i=59\degr$) the disk's angular momentum vectors would be $66\degr$ away from each other. Therefore, there is strong evidence suggesting that the SR24 system has two highly-misaligned disks. 

\subsection{Extended gas features}
\begin{figure*}[!h]
\epsscale{1.18}
\plotone{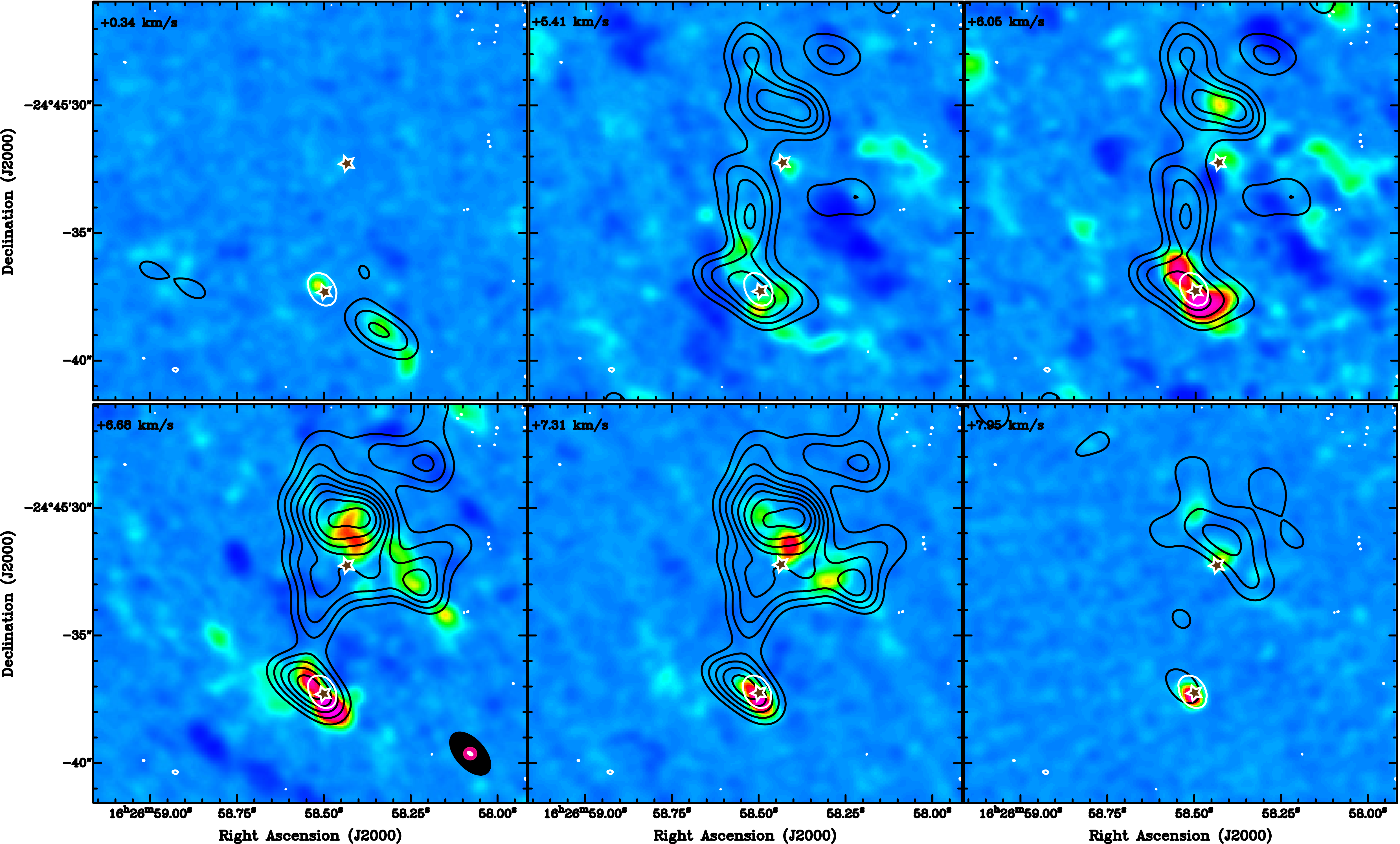}
\caption{\co(2-1) emission velocity channels observed with the SMA (black contours ranging from 0.22\jykms to 1.1\jykms in steps of 0.1\jykms) and 1.3~mm continuum emission from ALMA (white contours at 0.20\mjy) on top of the \co(2-1) emission velocity channels taken with ALMA (colour image) toward the SR24 system. The central velocity of the channels is shown at the top left corner of each panel. The synthesized beams are shown in the bottom left panel and share the same color code with the contours and stars mark the position of SR24N and SR24S}
\label{Fcube}
\end{figure*}

\begin{figure*}[!h]
\epsscale{1}
\plotone{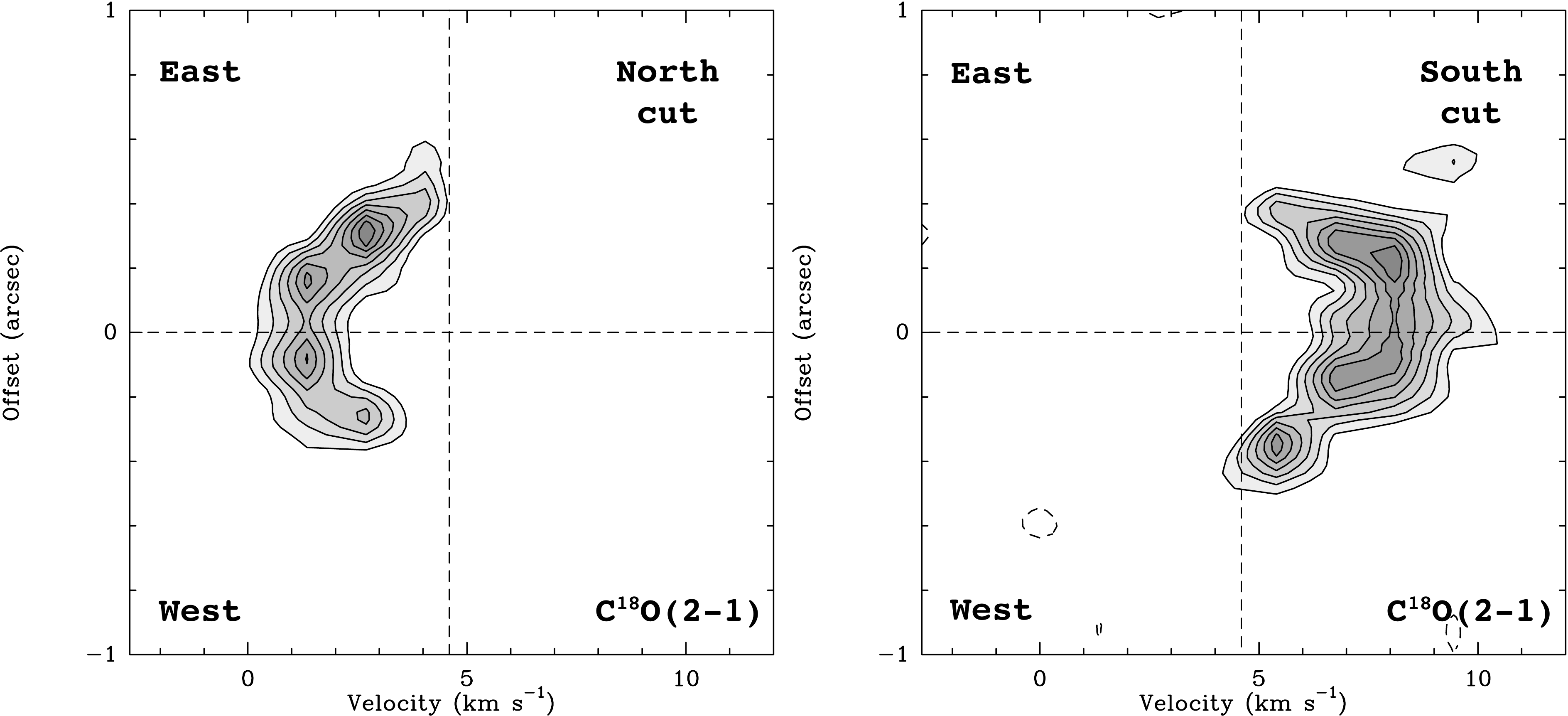}
\caption{SR24S position-velocity diagrams of the \ceto~(2-1) emission made with two cuts perpendicular to the disk's major axis (P.A.$=122\degr$). Velocity is in LSR and positional offsets are with respect to the major axis of the disk. Contours are at -20,20,30,40,50,60,70,80 and 90\% of the maximum peak emission. The left panel shows the cut taken North of the central position and the right panel shows the cut taken South of it. The asymmetric boomerang like features displayed are due to gas rotation and accretion.} 
\label{Fpvperp}
\end{figure*}

\begin{figure*}[!h]
\epsscale{1}
\plotone{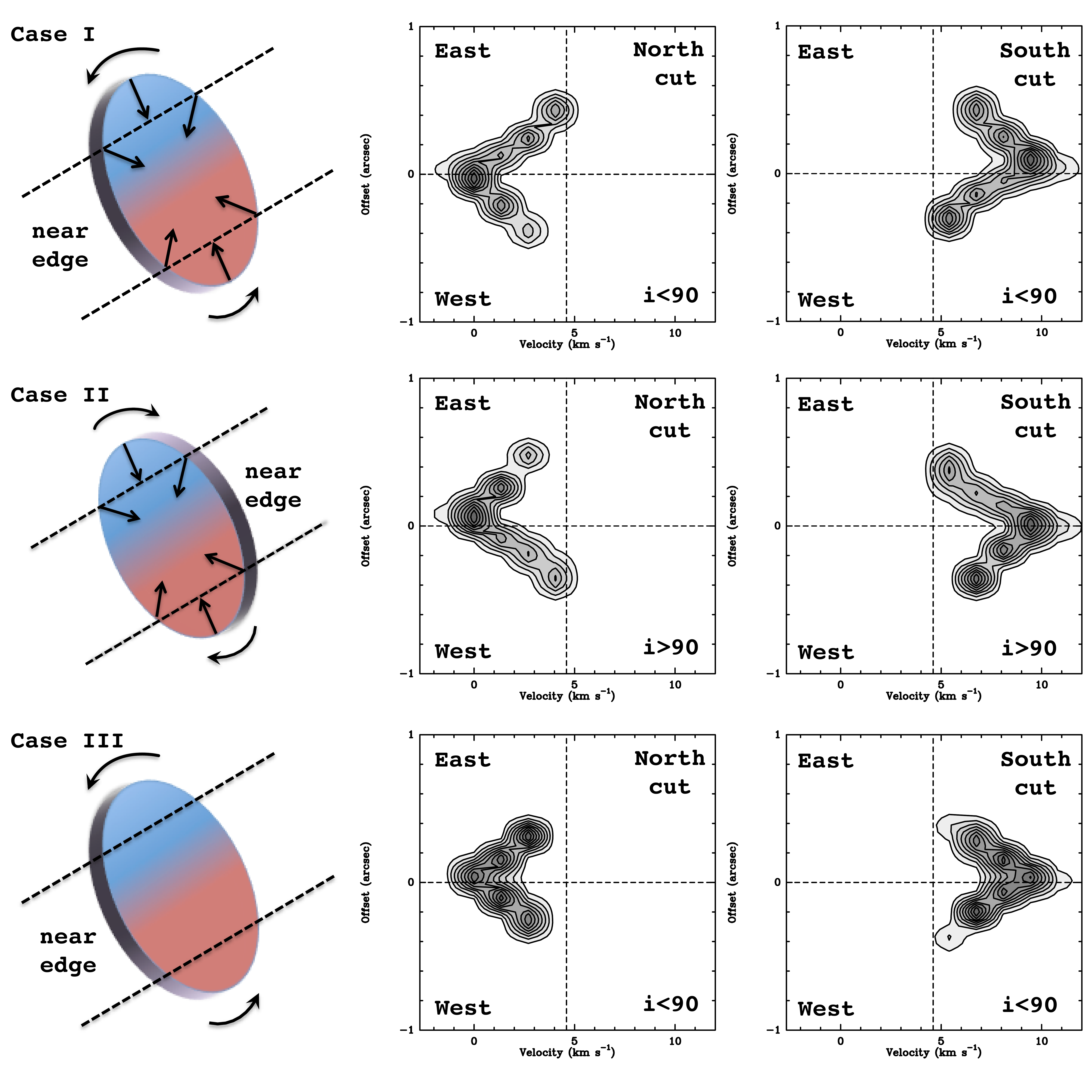}
\caption{Figure displays three cases of disks with different inclinations with respect to the plane of the sky and kinematics (sense of rotation and infall). We took the synthetic model for SR24S disk, changed some parameters and produced two associated position-velocity cuts perpendicular to its major axis (that is, with a P.A.=$122\degr$) and centered North and South of the disk center. The pv-diagrams are  displayed to the right of the corresponding disk sketch. In all three cases the blueshifted emission is produced in the NE and the redshifted emission is in the SW. In Case~I the disk has an inclination of $45\degr$ ($<90\degr$), the near edge due East, a counterclockwise rotation ($v_{rot}=3.2$\kms) and an infall velocity of 1.0\kms. In Case~II the disk inclination is $135\degr$ ($>90\degr$), the near edge is due West, it is rotating clockwise and has an infall velocity of 1.0\kms. Finally, Case~III shows the same disk as for Case~I but without infall velocity. For all three models $v_{LSR}=4.6$\kms.
}
\label{Fpvmodels}
\end{figure*}

\begin{figure*}[!h]
\epsscale{0.3}
\plotone{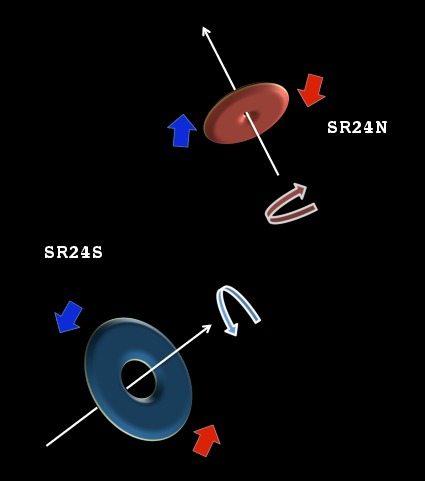}
\caption{Cartoon explaining the geometry and kinematics of the SR24 triple system. The picture is not intended to be a scaled version of the system, nor to display the morphological details of the disks. The inclinations and position angles are just roughly approximated.}
\label{Fcartoon}
\end{figure*}

Inspecting the HST optical emission of Figure \ref{Fcont}, it is noticeable the bridges and/or streamers produced by light scattered in dust and gas \citep{2003Potter}. There is a twisted bifurcation in the bridge joining the SR24N and SR24S disks, and a prominent curved arc south of SR24S thought to be a spiral arm \citep{2010Mayama}. The present \co (2-1) observations from ALMA and SMA do not perfectly match these structures. For instance, the southern optical/infrared spiral arm has no \co counterpart. The location of the molecular bridge between the two disks (Figure \ref{Fcube}) is slightly east of the HST bifurcated bridge, but both seem to join the two disks of the system. The nature of this bridge is unknown, although it is well known from numerical simulations that multiple systems have bridges of material like the one observed here \citep[e.g.,][]{2017Gabbasov}. We have also noticed that the optical bridge between the disks of the system has a curvature and joins the west side of SR24S with the east side of SR24N. Models including two disks with the same sense of rotation \citep[see Figure 1 in][]{2010Mayama} seem to predict a reversed orientation of the linking bridge and this is a point for further investigation of the geometry and kinematics of the system.
 
Regarding the northern gas reservoir seen in the \co (2-1) maps (Figure \ref{Fcube}), it is possible that it is connected (in space and velocity) with the bridge. It could be part of a spiral structure due to gravitational instabilities of the system \citep[see e.g.,][]{2016Tobin}, the remnant of a circumbinary structure or a reservoir of material found by SR24N in its orbital path. More sensitive ALMA observations are needed to go further interpreting the nature of this feature.

\subsection{Origin of the SR24 system}
The ALMA observations and the previous work on the SR24 system make it clear that the two disks are misaligned making this one more of a small sub-population of multiple systems with large separations \citep[][]{2014Jensen,2014Salyk,2014Williams}. However, SR24 has another peculiarity: the two disks possibly rotate in opposite directions as seen from Earth (i.e.; in projection, one rotates clockwise and the other counterclockwise). How this system has been originated is not a trivial question and here we pretend to introduce just some suggestions into the discussion.

It is known that a highly-misaligned multiple system is hardly formed via the collapse of a non-turbulent rotating cloud, and this has been previously stated by numerical simulations \citep[e.g.,][]{2010Bate}. Such non-turbulent clouds may produce disks that can undergo further fragmentation due to gravitational instabilities, being the most recent and clear example the triple system L1448 IRS3B \citep{2016Tobin}. Therefore a cloud would need some degree of initial turbulent motions as to fragment into (at least two) pieces that would finally rotate in highly misaligned, or even opposite directions \citep[i.e. antiparallel spins][]{2010Offner,2014Tokuda}. 

Another possibility to form a system with two strongly misaligned disks is the event of a close encounter of two previously isolated protostellar disks. However, in the thorough analysis by \cite{2014Salyk} adapting the expressions of \cite{2005Thies} to the Ophiuchus molecular cloud, they estimate a low encounter probability ($<10^{-5}$) for impact parameters under 500~AU. 
 
Whatever the multiple system formation mechanism was, at some point, one of the protostellar disk companions may undergo an event of external accretion from a reservoir of material \citep{2016Vorobyov}, that may impinge a tilt in this particular disk resulting in a tilt with respect to its original inclination. Regarding the existence of a gas reservoir north of SR24N, this may be a plausible possibility for the misalignment in SR24, which may erase the \textit{memory} of the formation mechanism in this system.

\section{Summary and Conclusions}
By using archival ALMA observations we report the detection at 1.3~mm of the two protostellar disks toward the SR24 system, both in continuum and molecular emission (\co and \ceto 2-1 lines). We estimate the disk masses for SR24S and SR24N (0.025\msun and $4\times10^{-5}$\mearth, respectively) and the dynamical masses for both protostars (1.5\msun and 1.1\msun). We implemented a simple kinematic model to fit the velocity cube of the SR24S \ceto disk emission and completely determine the geometry of the disks, and derive the possible sense of rotation and the angle between their angular momentum vectors. The disks are highly misaligned ($\sim108\degr$) and are probably rotating in opposite directions as seen from Earth.

We also use archival \co~(2-1) SMA observations to show the extended molecular emission distribution. We detect a bridge of gas joining both disks (other bridge is observed in optical/infrared as well) and a gas reservoir north of SR24N. The true nature of this gas reservoir may be the key for interpreting the formation of this system. If it is a remnant of a circumbinary disk, or part of a spiral structure associated with SR24N, then the large misalignment between the disks may suggest a turbulent origin of the system;  contrarily, if the gas reservoir is external to the system, it could well being accreted by SR24N, tilting its original orientation, and thus erasing all the information about the formation of this system (whether it was from the turbulent fragmentation of a cloud or the fragmentation of a disk).
 
\acknowledgments
We thank the anonymous referee for the thoughtful and dedicated 
comments and suggestions with which the manuscript substantially
improved.

This paper makes use of the following ALMA data: 2013.1.000912.S 
and 2013.1.00498.S. ALMA is a partnership of ESO
(representing its member states), NSF (USA) and NINS
(Japan), together with NRC (Canada) and NSC and
ASIAA (Taiwan) and KASI (Republic of Korea), in
cooperation with the Republic of Chile. The Joint
ALMA Observatory is operated by ESO, AUI/NRAO
and NAOJ.

The Submillimeter Array is a joint project between the 
Smithsonian Astrophysical Observatory and the Academia Sinica 
Institute of Astronomy and Astrophysics and is funded by 
the Smithsonian Institution and the Academia Sinica.

Facilities: \facility{ALMA, SMA}
\newpage
\bibliography{biblio}


\begin{deluxetable*}{cccccccc}
\tablewidth{0pt}
\tablecolumns{8}
\tablecaption{$^{12}$CO disks properties}
\phs
\tablehead{
\colhead{Column} & \colhead{2} & \colhead{3} & \colhead{4} & \colhead{5} & \colhead{6} & \colhead{7} & \colhead{8} \\
\tableline \\
\colhead{Source} & \colhead{Size} & \colhead{P.A.} & \colhead{i} & \colhead{V$_{center}$\tablenotemark{$\dagger$}} & \colhead{V$_{range}$\tablenotemark{$\dagger$}} & \colhead{V$_{peaks}$\tablenotemark{$\dagger$}} & \colhead{V$_{absorption}$\tablenotemark{$\dagger$}} \\
\colhead{}  & \colhead{(b$_{maj}\times$b$_{min}$;~P.A.)} & \colhead{($\degr$)} & \colhead{($\degr$)} & \colhead{(km~s$^{-1}$)}  & \colhead{(km~s$^{-1}$)} & \colhead{(km~s$^{-1}$)} & \colhead{(km~s$^{-1}$)} 
}  
\startdata
SR24S & 0\farcs50$\pm$0\farcs06$\times$0\farcs17$\pm$0\farcs02; ~41$\degr$ & 218$\pm$2  & 70$\pm5~$ &  4.4$\pm$0.3 & [-3.5,12.4] & -0.3/6.0 & [2.2,4.1] \\
SR24N & 0\farcs60$\pm$0\farcs10$\times$0\farcs31$\pm$0\farcs06; 111$\degr$ & 297$\pm$5 & 121$\pm17$ & 5.6$\pm$0.3 & [-0.3,10.5] &  ~1.6/7.3 & [2.2,4.8] \\
\enddata 
\tablecomments{\textbf{Column 2:} Deconvolved FWHM of the semimajor and semiminor axis and position angle estimated from a 2D-Gaussian fit to the moment 0 image produced using the V$_{range}$ specified in column 5. \textbf{Column 3:} Estimated using a linear fit to the peak of the red- and blue-shifted high-velocity channels. \textbf{Column 4:} Derived from the semimajor and minor axis obtained in column 2. \textbf{Column 5:} Measured from a Gaussian fit to the wings of the line profile integrated along the disk. \textbf{Column 6:} Measured until no 3$\sigma$ emission is detected in a velocity channel at the disk position. \textbf{Column 7:} Velocities of the disk's double peak spectrum. \textbf{Column 8:} Velocity range of the self-absorption feature.}
\tablenotetext{$\dagger$}{Velocities are $V_{LSR}$ and are not corrected by the cloud velocity.}
\label{T12co}
\end{deluxetable*}

\begin{deluxetable*}{lr@{$\pm$}l}
\tablewidth{0pt}
\tablecolumns{3}
\tablecaption{C$^{18}$O fit disk parameters}
\phs
\tablehead{
\colhead{Parameter} & \multicolumn{2}{c}{Fit Value}
}
\startdata
$x_0$ \tablenotemark{$\dagger$}    &     $0\farcs02$ & $0\farcs02$\\
$y_0$ \tablenotemark{$\dagger$}    &     $-1\farcs30$ & $0\farcs03$\\
$i$          &     $46\degr$ & $8\degr$\\
$r_c$      &     $1\farcs3$ & $0\farcs2$\\
$p.a.$     &     $212\degr$ & $6\degr$\\
$v_{LSR}$\tablenotemark{$\star$}  & 4.6 & 0.2 \\
$v_{rot}$\tablenotemark{$\star$}   & 3.2 & 0.5 \\
$v_{infall}$\tablenotemark{$\star$} & 0.3 & 0.2 \\
$F_0\times10^{-5}$ &   4 & 1\\
\enddata 
\tablecomments{Results from the \ceto velocity cube fitting to the SR24S disk.}
\tablenotetext{$\dagger$}{These offsets set the modeled disk center at 16:26:58.507,-24:45.37.17.}
\tablenotetext{$\star$}{Velocities are in \kms.}
\label{Tc18o}
\end{deluxetable*}

\begin{deluxetable}{cccc}
\tablewidth{0pt}
\tablecolumns{4}
\tablecaption{SMA $^{12}$CO extended emission}
\phs
\tablehead{
\colhead{Source} & \colhead{Flux Density} & \colhead{Size} & \colhead{M$_{H_2}$} \\
\colhead{}  & \colhead{(Jy~km~s$^{-1}$)} & \colhead{(arcsec$^2$)} & \colhead{($\times10^{-7}$~M$_{\sun}$)}  
}
\startdata
Northern reservoir  & 11.3$\pm$0.3 & 16.5 & 13 \\
Bridge                   & 3.0$\pm$0.1 & 37.5 & 3 \\
Southern streamer & 1.6$\pm$0.1 & 7.0 & 2 \\
\enddata 
\tablecomments{\textbf{Column 2:} Integrated \co(2-1) flux densities from SMA observations with a synthesized beam of $2\farcs0\times1\farcs5$; P.A.$=42\fdg2$. \textbf{Column 3:} Approximated size of the source taken from the SMA dataset. \textbf{Column 4:} Mass of the gas estimated from the SMA integrated flux measurements.}
\label{Textended}
\end{deluxetable}


\end{document}